**Topological Structure of Urban Street Networks from the Perspective of Degree Correlations**

Bin Jiang[1], Yingying Duan[2], Feng Lu[2], Tinghong Yang[3] and Jing Zhao[3]

[1]Department of Technology and Built Environment, Division of Geomatics
University of Gävle, SE-801 76 Gävle, Sweden
Email: bin.jiang@hig.se

[2]State Key Laboratory of Resources and Environmental Information System
Chinese Academy of Sciences, 11A Datun Road, Beijing 100101, China
Email: duanyy37@hotmail.com, luf@lreis.ac.cn

[3]Department of Mathematics, Logistical Engineering University,
Chongqing 400016, China
Email: yth_ah@163.com, zhaojanne@gmail.com

*(Draft: August 2011, Revision: July 2012, February, August 2013)*

**Abstract**
Many complex networks demonstrate a phenomenon of striking degree correlations, i.e., a node tends to link to other nodes with similar (or dissimilar) degrees. From the perspective of degree correlations, this paper attempts to characterize topological structures of urban street networks. We adopted six urban street networks (three European and three North American), and converted them into network topologies in which nodes and edges respectively represent individual streets and street intersections, and compared the network topologies to three reference network topologies (biological, technological, and social). The urban street network topologies (with the exception of Manhattan) showed a consistent pattern that distinctly differs from the three reference networks. The topologies of urban street networks lack striking degree correlations in general. Through reshuffling the network topologies towards for example maximum or minimum degree correlations while retaining the initial degree distributions, we found that all the surrogate topologies of the urban street networks, as well as the reference ones, tended to deviate from small world properties. This implies that the initial degree correlations do not have any positive or negative effect on the networks' performance or functions.

**1. Introduction**
Many complex networks demonstrate a striking degree correlation, i.e., a node tends to link to other nodes with similar (or dissimilar) degrees. This property of degree correlation has been well-studied in the literature to characterize different real-world networks (Newman 2002, 2003, Serrano et al. 2007). For example, many social networks demonstrate a positive degree correlation (a high probability of linking to similar degrees). Other networks, such as biological and technological networks, show a negative degree correlation (a high probability of linking to dissimilar degrees). This sounds very much like the phenomenon of spatial autocorrelation, which was formulated as the first law of geography: "*Everything is related everything else, but near things are more related than distant things*" (Tobler 1970, p. 236, Goodchild 1987). This degree correlation can also be characterized by the rich-club phenomenon (Zhou and Mondragón 2004), i.e., a highly connected node tends to get connected with other highly connected nodes. From the perspective of degree correlations, this paper attempts to examine the topological structure of various urban street networks in comparison with three reference networks (biological, technological, and social).

Urban street networks studied in this paper are the topologies or relationships of individual streets rather than those of the street segments. Such a network topology is a de facto graph $G = (V, E)$, consisting of vertices representing the individual streets and edges if the corresponding streets intersect. This network topology describes the information aspect of the street network by indicating the way a person navigates the city (Rosvall et. al. 2005, Masucci et al. 2009), so it has many modeling advantages compared to the alternative network representation of street segments or junctions (Jiang and Claramunt 2004, Rosvall et al. 2005); the network topology helps to address various issues that can hardly be addressed by the alternative representation. For example: Which streets are prominent or marginal, according to the overall structure of the underlying network topology? How many intermediate streets must one pass to reach one from another?

The alternative network representation of relationships between street junctions is essentially geometric in the sense of junction locations and segment distances. The duality of this network can be the relationship of street segments, but the essential geometric properties remain. This geometric network can be used for computing distances, routing, and tracking, whereas the network topology can uncover underlying structures or patterns. More importantly, the network topology is closer to the human conception of street space in terms of how individual streets intersect with each other. Because of these advantages, the network topology or topological network has emerged as an important means to study urban forms and patterns from the perspective of complex networks (e.g., Jiang and Claramunt 2004, Porta et al. 2006, Jiang 2007). For example, previous studies have shown that the topological networks exhibit small-world and scale-free properties.

The current study illustrated some interesting topological properties of urban street networks through comparison with the reference networks based on degree correlations and through reshuffling the topologies of the complex networks towards for example maximum or minimum degree correlations. This paper examined topological properties using multiple measures of degree correlations from both local and global perspectives. Unlike the three reference networks, urban streets topologies show little degree correlations. Yet almost all the urban street networks show some consistent topological structures for either European or North American networks. We concluded that the initial degree correlations do not imply high efficiency or high performance of the networks.

The remainder of this paper is organized as follows. Section 2 presents the data and methods used in the study, including six urban street networks and three reference networks. Section 3 shows two major results: (1) a lack of degree correlations for urban street network topologies in general, and (2) no relationship between degree correlations and network's efficiency. Section 4 further draws on related evidence in the literature to support our findings, and explains why our investigation is more robust. Finally Section 5 draws a conclusion and points to our future work.

## 2. Data and methods
This section presents the six urban street networks and three reference networks respectively taken from an open source and the literature. To characterize topological structure of urban street networks or complex networks in general, we introduce a series of measures about degree correlations, network efficiencies, and network reshuffling processes.

### 2.1 Data collection and network construction
We adopted six street networks from Copenhagen, London, Paris, Manhattan, San Francisco, and Toronto. They represent some typical street patterns in the literature (Jacobs 1995, Figure 1); Unlike the European networks, the North American networks were more grid-like, and the streets were mostly straight. The data of the street networks was taken from OpenStreetMap database, a free editable map of the world (Bennett 2010). The initial OSM data are just individual lines, and there is no information about which line intersected other lines. To derive this information, we must build up topology (Note: this topology is different from the topology referred to in this paper), which is an ordinary vector GIS processing. This processing essentially establishes a geometric network, showing how individual junctions are connected via street segments. After building up the topology, each segment becomes directed so that left and right polygons can easily be identified. Based on the geometric network, the every-best-fit join principle, which shares the same spirit of the intersection continuity negotiation principle (Porta et al. 2006), was applied to merge those adjacent segments to form individual natural streets (Jiang et al. 2008). Topologies of the street networks were set up by taking individual streets as the nodes and street intersections as the edges of a graph. The street network topologies were further investigated and compared to the three reference networks.

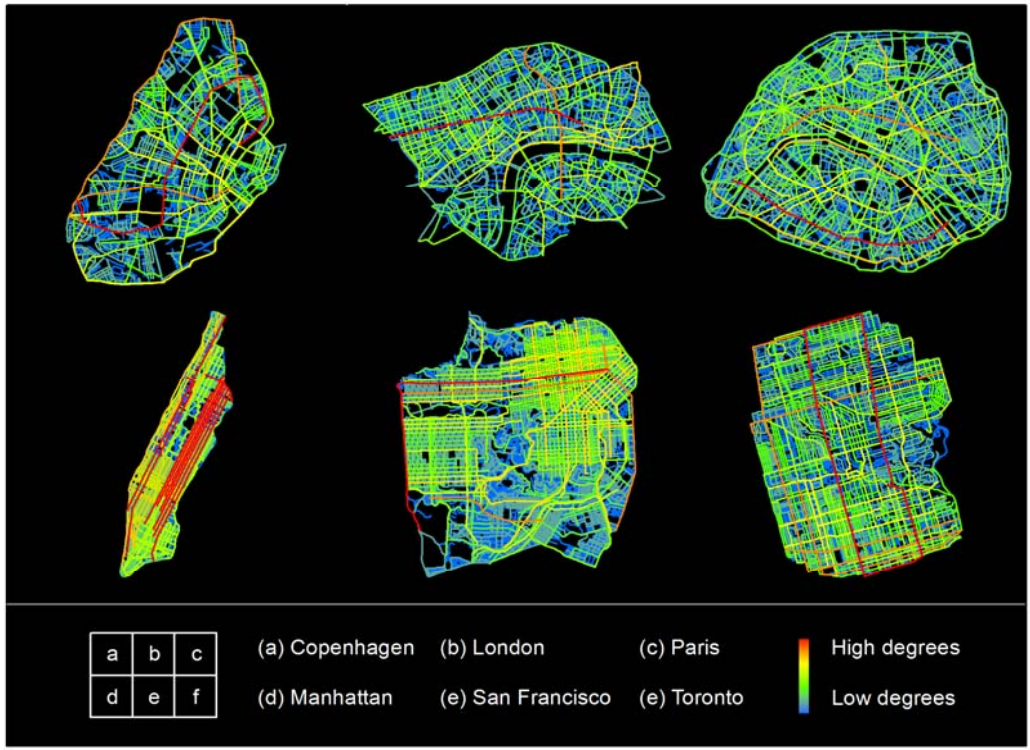

Figure 1: (Color online) Six urban street networks (three European and three North American)

(Note: The coloring is based on the head/tail breaks in order to reveal the underling scaling pattern (Jiang 2013); this is the same for Figures 3 and 4)

The three reference networks are respectively biological, technological, and social, and they are protein interactions (Maslov and Sneppen 2002, Colizza et al. 2005), the Internet (e.g., Pastor-Satorras and Vespignani 2004, Faloutsos et al. 1999), and scientific collaboration (Newman 2001a, 2001b). This paper used the same network topologies that Zhou and Mondragón (2007) previously studied. Nine complex networks (six urban street network topologies, and three other networks) were studied in this paper. All nine networks had similar sizes of approximately hundreds to thousands of vertices. Please refer to Table 1 in Section 3.1 for more detailed information about the networks.

**2.2 Degree correlations and related measures**
Inspired by the work of Barabási and Albert (1999) for generating scale-free networks using a preferential attachment mechanism, Newman (2002) observed that this preference often occurred among nodes with similar or dissimilar degrees. In other words, the preferential attachment mechanism considers only target nodes (for example, the richest nodes or highly connected ones) for a link. The concept of degree correlation suggests that both source and target nodes should be considered for preferential attachment, i.e., similar or dissimilar nodes tend to link together. Newman (2002, 2003) suggested an assortativity coefficient to measure the degree correlation based on the Pearson correlation coefficient of the degrees at either ends of an edge. The assortativity coefficient is formally defined as:

$$A = \frac{e^{-1}\sum_{j>i}a_{ij}d_id_j - [e^{-1}\sum_{j>i}a_{ij}\frac{d_i+d_j}{2}]^2}{e^{-1}\sum_{j>i}a_{ij}\frac{d_i^2+d_j^2}{2} - [e^{-1}\sum_{j>i}a_{ij}\frac{d_i+d_j}{2}]^2},  \quad [1]$$

where $e$ is the number of edges, $d_i$ and $d_j$ are the degrees of nodes $i$ and $j$, $a_{ij}$ is an element of the network or graph's adjacency matrix, $a_{ij}=1$, if nodes $i$ and $j$ are linked each other. Otherwise, $a_{ij}=0$. $A>0$ means that a network is assortative mixing, and $A < 0$ means disassortative mixing.

An alternative measure of degree correlations is to check the correlation between the nodes' degrees and the average degree of its neighbors, known as k nearest-neighbor connectivity (Pastor-Satorras et al. 2001). If the correlation or slope is positive, the network is assortative or holds positive degree correlations. If the correlation or slope is negative, the network holds disassortative mixing or negative degree correlations. K nearest-neighbor connectivity is defined by

$$K = \sum_{d_j} d_j P_c(d_j | d_i), \quad [2]$$

where $P_c(d_j | d_i)$ is the conditional probability that an edge of node degree $d_i$ points to a node with degree $d_j$. Unlike assortativity coefficient that is a value between -1 and +1, the $K$ measure is through the slope of the degree-degree plot to indicate degree correlations.

On the other hand, rich-club coefficient can also reflect the degree correlations in a network from another perspective. The rich-club coefficient is defined by the ratio of the edges or links among the rich nodes to the possible edges for the rich nodes, forming a complete graph (Zhou and Mondragón, 2004). The rich-club coefficient is expressed by

$$\phi(k) = \frac{2e_{>k}}{n_{>k}(n_{>k}-1)}, \quad [3]$$

where $e_{>k}$ is the number of edges among the rich nodes – the nodes with degrees greater than $k$., $n_{>k}$ denotes the number of rich nodes with degrees greater than $k$. In fact, $\frac{n_{>k}(n_{>k}-1)}{2}$ represents the possible number of edges for the rich nodes, forming a complete graph.

With respect to Equation 3, the concept of rich nodes is relative. Given different $k$ values, there are a series of rich-club coefficients. This is clearly different from assortativity coefficient, with one value for characterizing the assortativity of the entire network topology. The rich-club coefficient needs a series of $\phi(k)$ values with respect to the different $k$ values to characterize the degree correlations. The rich-club coefficient of a complex network should be compared to its random counterpart to fully characterize degree correlations.

Since high-degree nodes have a larger probability of sharing edges than low-degree nodes, $\phi(k)$ is usually a monotonic increasing function of $k$ in many complex networks, including the Erdos-Renyi random network. Thus, only the monotonically increasing feature of $\phi(k)$ will not be a signature of the rich club phenomenon. The normalization of $\phi(k)$ is defined as follows (Colizza et al. 2006):

$$R_{rand}(k) = \frac{\phi(k)}{\phi_{rand}(k)}, \quad [4]$$

where $\phi_{rand}(k)$ is the rich-club coefficient of the maximally random network with the same degree distribution *P(k)* of the network. An actual rich club ordering is denoted by ratio $R_{rand}(k) > 1$. If $R_{rand}(k) < 1$ there is an anti-rich club ordering, while $R_{rand}(k) = 1$ suggests an absence of rich club ordering. This study constructed 500 maximally random rewiring networks for each network, and then used their average rich-club coefficient as $\phi_{rand}(k)$ in Equation 4.

## 2.3 Characterizing network's small world properties
A small world network can be characterized at both local and global scales and using clustering coefficient and average path length, respectively. The clustering coefficient *C(v_i)* of node *v_i* is defined as the ratio of the actual number of edges to the possible number of edges among neighbors of node *v_i* (Watts and Strogatz 1998). The clustering coefficient of a network is defined as the average of *C(v_i)* over all *v_i*:

$$C = \frac{1}{n}\sum_{i=1}^{n}\frac{2|N(v_i)|}{d(v_i)(d(v_i)-1)}, \quad [5]$$

where $|N(v)|$ denotes the number of edges between neighbors of node *v_i* and *d(v_i)* is the degree of *v_i*.

The average length of the shortest paths between any pair of nodes in the network:

$$L = \sum_{i=1}^{n}\sum_{j=1}^{n} s_{ij} / n(n-1), \quad [6]$$

where $s_{ij}$ is the shortest topological distance between a pair of nodes i and j.

The average path length indicates a sort of compactness for a network topology through the shortest topological distance for all pairs of the nodes. At the global scale, if two randomly picked nodes are separated by a small number of intermediate nodes, the network is considered to be more efficient. The shorter the average path length is, the more efficient a network is at the global scale. This insight about network efficiency was first captured by Latora and Marchiori (2001) with a particular metric for measuring the efficiency, i.e., small world networks are capable of efficiently transferring information locally and globally. In this paper, we will adopt the above two small world properties to examine how a surrogated network deviates from a small world, and thus become less efficient.

## 2.4 Network reshuffling processes
The links of the original networks were reshuffled in four different ways to better understand the influence of degree correlations on the network's performance and efficiency: maximum assortativity (correlation+), minimum assortativity (correlation-), random and neutral rewiring (Holme and Zhao 2007, Zhao et al. 2007). For the random rewiring or reshuffling, in each step a pair of edges was randomly chosen and rewired. This reshuffling was done only if it generated no multiple edges while keeping the graph connected. The reshuffling process kept the degree of each node unchanged. Repeating this process for a sufficient number of times generates a maximally random rewiring of the original network. Each step only accepted changes moving toward the desired direction to get the three types of reshuffling networks. More specifically, given $\Delta A = A_{i+1} - A_i$, if the rewiring process was executed, it was only conducted to make $\Delta A > 0$ for maximization of *A* (correlation+). The opposite was done for the minimization of *A* (correlation-). Similarly, the reshuffling process was repeated many times until the assortativity coefficient converged to a stable value, such as $\Delta A < 0.0001$, where *i* denotes the iteration number. The last reshuffling was neutral, in which only links between the richest and the second poorest, and the second richest and the poorest, were allowed. Figure 2 provides an illustration for the last three reshuffling processes.

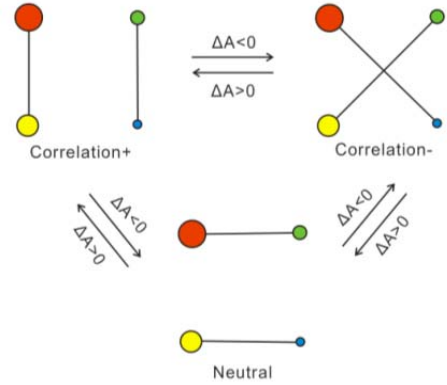

Figure 2: (Color online) Illustration of the three different reshuffling processes (red = richest node, yellow = second richest node, blue = poorest node, green = second poorest node)

## 3. Results and discussion

Having introduced the data and methods, this section shows the results developed from the experiments. This section first examines some basic facts about the networks. Following the network construction, six network topologies were obtained. Together with the three reference networks, basic statistics of the set of complex networks in terms of nodes, edges, average degrees, degree distribution, and their small world properties (characterized by clustering coefficient and path length) are shown in Table 1. All the complex networks possess scale-free or small-world properties. A very robust method based on maximum likelihood estimation (Clauset et al. 2009) was used to detect scale-free properties, so it was very likely that the power law exponents computed for the three reference networks differ from the previous studies. The column P indicates the goodness of fit. The higher the P values, the better fit for degree distributions to a power law. Usually $P > 0.01$ or $0.05$ is acceptable. Table 1 indicates that all nine networks exhibit very striking power law distributions.

Table 1: Some facts about the nine network topologies
(N = Nodes, E = Edges, Mean(d) = Average degree, P(d) = Degree distribution, P = Goodness of fit for a power law, C = Clustering coefficient, L = Path length)

|   | N | E | Mean(d) | P(d) | P | C | L |
|---|---|---|---|---|---|---|---|
| Copenhagen | 1637 | 3410 | 4.2 | $P(d) \sim d^{-2.6}$ | 0.09 | 0.28 | 4.52 |
| London | 3010 | 6287 | 4.2 | $P(d) \sim d^{-2.6}$ | 0.06 | 0.26 | 5.50 |
| Paris | 4501 | 11408 | 5.1 | $P(d) \sim d^{-2.3}$ | 0.07 | 0.37 | 5.36 |
| Manhattan | 1046 | 4617 | 8.8 | $P(d) \sim d^{-2.4}$ | 0.12 | 0.33 | 3.79 |
| San Francisco | 3110 | 9999 | 6.4 | $P(d) \sim d^{-2.3}$ | 0.13 | 0.31 | 4.88 |
| Toronto | 2599 | 5904 | 4.5 | $P(d) \sim d^{-2.5}$ | 0.11 | 0.27 | 4.58 |
| Protein | 4626 | 14801 | 6.4 | $P(d) \sim d^{-3.1}$ | 0.16 | 0.12 | 4.22 |
| Internet | 9200 | 28957 | 6.3 | $P(d) \sim d^{-2.2}$ | 0.04 | 0.62 | 3.12 |
| Scientific | 12722 | 39967 | 6.3 | $P(d) \sim d^{-3.5}$ | 0.08 | 0.72 | 6.83 |

### 3.1 Visualization of Network topologies

The nine network topologies are first visualized using the graph exploration system GUESS (Adar 2006) through the graph layout based on a generalized expectation maximization algorithm (Dempster et al. 1977). This layout has a good visual effect because of minimum edge intersections imposed by the particular algorithm. This paper uses the same layout for visualization of network topologies, so the different visual effects indicate potential differences between the network topologies. Figure 3 illustrates the network topologies in which both node sizes and colors indicate the difference of degrees for individual nodes. It is difficult, just based on the visualization, to assess how the street network topologies differ from the three reference network topologies.

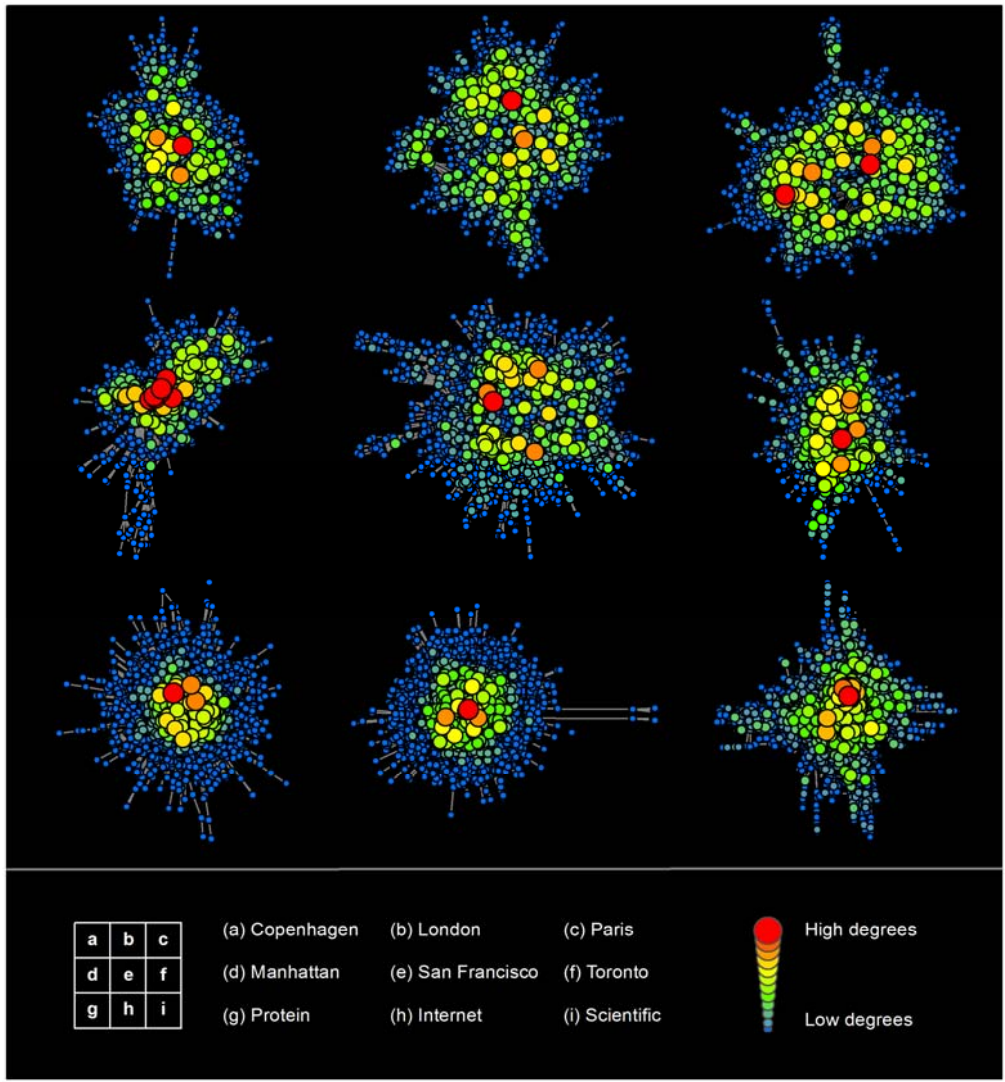

Figure 3: (Color online) Visualization of nine network topologies with minimum edge intersections

The graph layout differentiates the four reshuffled statuses: maximum assortativity (correlation+), minimum assortativity (correlation-), random, and neutral. Taking the Copenhagen network topology for example, the four reshuffled statuses are very distinct (Figure 4). This is particular true for the maximum assortativity and minimum assortativity statuses. The former looks like an octopus, while the latter looks like a shoe. The two extreme assortativity cases clearly differ from random and neutral cases.

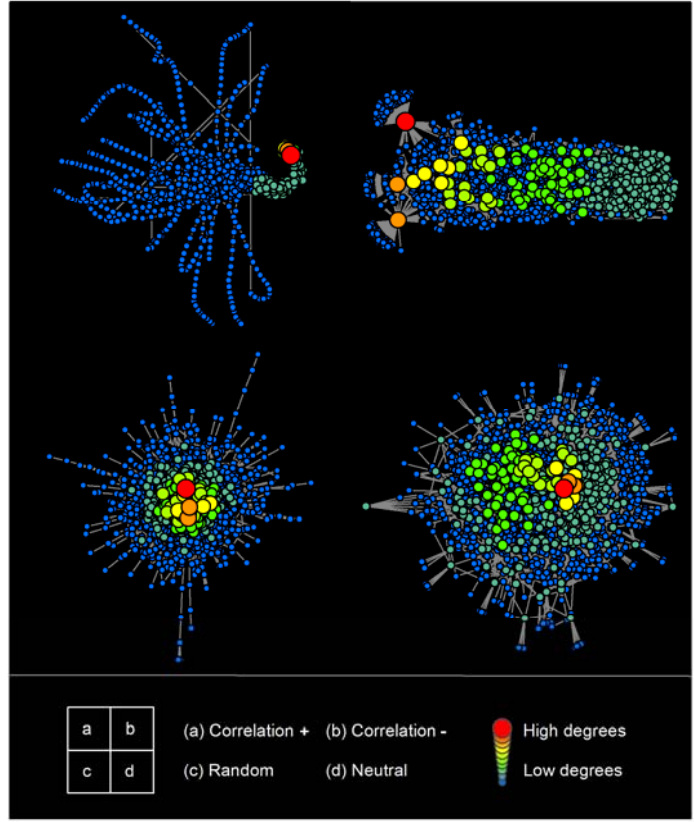

Figure 4: (Color online) Four network topologies derived from the Copenhagen example through the reshuffling processes

### 3.2 Lack of degree correlations for the street network topologies

In order to obtain insights into the degree correlation profiles of the networks, we computed $A$ for the nine networks and their reshuffled networks (Table 2 and Figure 5). The street network topologies (except Manhattan) demonstrate a consistent pattern, i.e., the original networks have assortativity coefficient very close to zero, and no statistically significant difference from their corresponding random counterparts (absolute values of the z-scores < 5). In contrast, the Manhattan network and the three reference networks demonstrate very strong degree correlations (either positive or negative), which differ significantly from their random counterparts (absolute values of z-scores > 15). In this regard, the Manhattan network clearly differs from the other five street networks. This was mainly due to the stripe-shaped geographic area of Manhattan. All major and long streets were along the length of the island, while the shortest streets were along the width of the island. This particular street layout creates the effect of rich-poor linkage. This is why the original Manhattan network topology shows a significant negative degree correlation. It should be noted that this peculiarity of the Manhattan street network was very unique. Most urban street networks have no such geographic constraint.

Table 2 Assortativity coefficients for the nine networks and their reshuffled counterparts
(Note: For the sake of accuracy, three decimal places are used for the mean values of the random reshuffling, and four decimal places for standard deviation [SD])

|  | Copenhagen | London | Paris | Manhattan | San Francisco | Toronto | Protein | Internet | Scientific |
|---|---|---|---|---|---|---|---|---|---|
| Original | -0.07 | -0.06 | -0.06 | -0.26 | -0.01 | -0.06 | -0.14 | -0.24 | 0.16 |
| Correlation+ | 0.25 | 0.61 | 0.71 | 0.13 | 0.84 | 0.15 | 0.34 | -0.22 | 0.92 |
| Correlation- | -0.28 | -0.37 | -0.43 | -0.50 | -0.66 | -0.21 | -0.31 | -0.28 | -0.55 |
| Neutral | -0.18 | -0.24 | -0.27 | -0.33 | -0.41 | -0.14 | -0.15 | -0.25 | -0.24 |
| Random | -0.060 | -0.025 | -0.020 | -0.157 | -0.033 | -0.058 | -0.033 | -0.236 | -0.002 |
| SD (random) | 0.0079 | 0.0101 | 0.0083 | 0.0065 | 0.0087 | 0.0049 | 0.0047 | 0.0002 | 0.0048 |
| Z-scores (original) | -1.30 | -3.49 | -4.78 | -15.86 | 2.58 | -0.37 | -22.83 | -17.18 | 34.10 |

The range of *A* varies from one network to another, but in general the ranges are generally much less than the theoretical limit [-1, +1]. This is clearly shown in Figure 5, where the Internet network has the shortest range, while the San Francisco and scientific networks have relatively wide ranges. In contrast to the six street network topologies, the three reference networks demonstrate very strong degree correlations, either positive or negative.

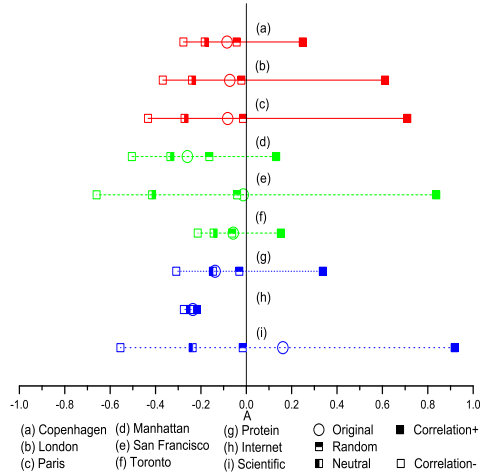

Figure 5: (Color online) Assortativity coefficient for the nine network topologies and their reshuffled statuses (Red = European networks, Green = North American networks, Blue = reference networks)

To confirm the above result about the absence of degree correlations of urban street networks, k nearest-neighbor connectivity was computed and plotted (Figure 6). The *k* nearest-neighbor connectivity clearly shows the absence of degree correlations for the street network topologies (Figure 6a and 6b). However, there were strong degree correlations for the three reference networks (Figure 6c). The absence of degree correlations for street network topologies is mainly due to the fact that on the one hand highly connected streets tend to connect to other highly connected streets, and on the other hand they also tend to connect less-connected streets as well. Eventually, the street network topologies overall lack of degree correlations. It should be noted that the significant negative correlation for Manhattan is was not obvious with the plot.

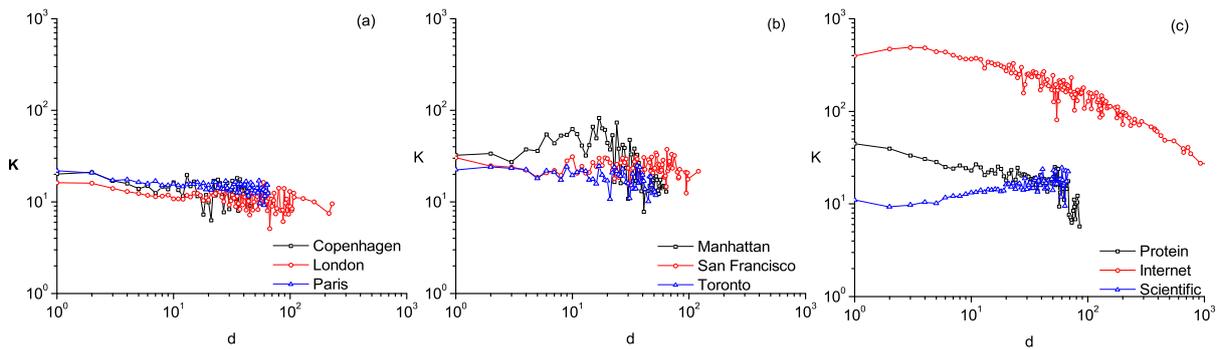

Figure 6: (Color online) *K* nearest-neighbor connectivity for (a) European networks, (b) North American networks and (c) reference networks
(Note: k and d indicate respectively k nearest-neighbor connectivity and node degree as shown in Equation 2.)

To further confirm the above result about the absence of degree correlations of urban street networks. Figure 7 shows the rich-club ratios $R_{ran}(k)$ for the nine networks. In the figure, the ratio $R_{ran} = \phi/\phi_{ran}$ was plotted as a function of the degree *k* and compared with the baseline value equal to 1 (the red lines). If $R_{ran}(k) > 1$, the network shows the presence of the rich-club phenomenon with respect to the random contrast networks. Conversely, if $R_{ran}(k) < 1$, the network shows anti-rich-club phenomenon. For the four networks with strong degree correlations (Manhattan and the three reference networks), the rich-club features are consistent with their assortativities. The scientific network, which was assortative, exhibits significant rich-club phenomenon, while the other three disassortative networks show significant anti-rich-club effect. However, the five street networks that lack significant degree correlation exhibit different profiles of rich-club phenomenon. Only Copenhagen showed rich-club features, while only the several richest nodes of London and San Francisco link in a rich-club way. Paris exhibits weak anti-rich-club ordering, while Toronto could be regarded as absence of rich-club ordering. This confirms that the rich-club phenomenon is not necessarily associated with assortative mixing, especially in networks that lack a significant degree correlation.

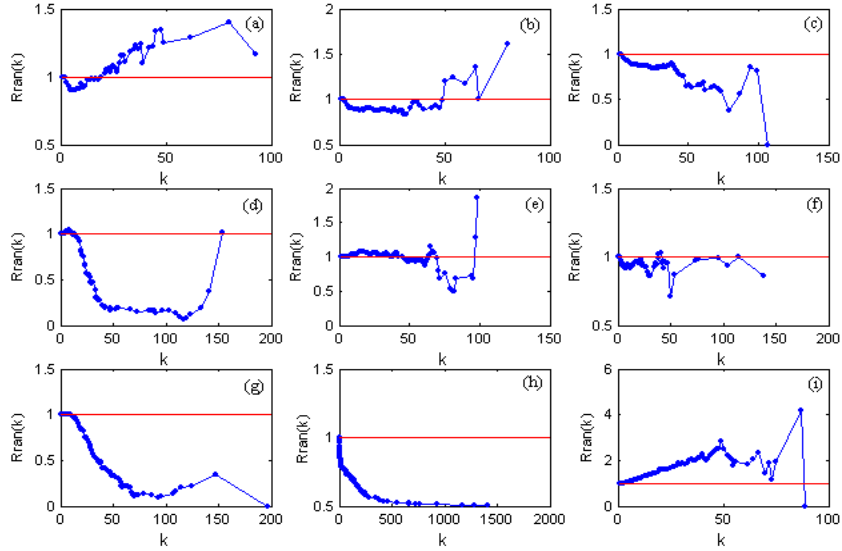

Figure 7: (Color online) Rich-club phenomenon for: (a) Copenhagen, (b) London, (c) Paris, (d) Manhattan, (e) San Francisco, (f) Toronto, (g) Protein, (h) Internet, and (i) Scientific

**3.3 Implications of degree correlations to network's small world properties**
We further examined the relationship between degree correlations and network's small world properties, and found that the original networks have the striking small world properties compared to all other statuses achieved through the reshuffling processes. The small world properties were measured by clustering coefficient at a local level and by the average path length at the global level. Table 3 shows that the original networks have significantly larger clustering coefficients than all of the counterparts. Their average path lengths are close to those of their random networks. This suggests that the original networks have the highest local efficiency compared to all other statuses achieved through the reshuffling process and almost the same global efficiency as their random contrast networks. The assortativity coefficient *A* of all street network topologies (with the exception of Manhattan) is very close to zero, and have no statistically significant difference from the corresponding random counterparts. However, the fact that the original network topologies have a much higher efficiency than the random ones implies that topologies with a similar assortativity coefficient may have distinct structural and functional differences. This finding about the relationship between degree correlations and network efficiency is applicable for all the nine network topologies. Table 3 lists assortativity coefficient *A*, clustering coefficient *C*, and average path length *L* of the nine networks and their contrast networks: maximum assortativity (correlation+), minimum assortativity (correlation-), and random and neutral networks (random and neutral).

Table 3: Small world properties with respect to the five statuses of the network topologies (A = Assortativity coefficient, C = Clustering coefficient, L = Average path length. For the sake of accuracy, some values kept to three decimal places)

|  |  | Original | Correlation+ | Correlation- | Random | Neutral |
|---|---|---|---|---|---|---|
| Copenhagen | A | **-0.07** | 0.25 | -0.28 | -0.06 | -0.18 |
|  | C | **0.28** | 0.07 | 0.00 | 0.03 | 0.00 |
|  | L | **4.52** | 32.72 | 7.55 | 4.21 | 6.68 |
| London | A | **-0.06** | 0.61 | -0.37 | -0.025 | -0.24 |
|  | C | **0.26** | 0.05 | 0.00 | 0.01 | 0.00 |
|  | L | **5.50** | 26.24 | 7.78 | 4.61 | 8.23 |
| Paris | A | **-0.06** | 0.71 | -0.43 | -0.02 | -0.27 |
|  | C | **0.37** | 0.04 | 0.00 | 0.01 | 0.00 |
|  | L | **5.36** | 31.90 | 8.02 | 4.29 | 7.61 |
| Manhattan | A | **-0.26** | 0.13 | -0.50 | -0.157 | -0.33 |
|  | C | **0.33** | 0.16 | 0.01 | 0.13 | 0.00 |
|  | L | **3.79** | 14.09 | 6.45 | 2.98 | 4.38 |
| San Francisco | A | **-0.01** | 0.84 | -0.66 | -0.033 | -0.41 |
|  | C | **0.31** | 0.07 | 0.00 | 0.02 | 0.00 |
|  | L | **4.88** | 19.77 | 8.78 | 3.85 | 6.94 |
| Toronto | A | **-0.06** | 0.15 | -0.21 | -0.058 | -0.14 |
|  | C | **0.27** | 0.06 | 0.00 | 0.02 | 0.00 |
|  | L | **4.58** | 24.08 | 7.46 | 4.16 | 6.35 |
| Protein | A | **-0.14** | 0.34 | -0.31 | -0.033 | -0.15 |
|  | C | **0.12** | 0.07 | 0.00 | 0.02 | 0.00 |
|  | L | **4.22** | 17.42 | 9.22 | 4.16 | 6.88 |
| Internet | A | **-0.24** | -0.22 | -0.28 | -0.236 | -0.25 |
|  | C | **0.62** | 0.19 | 0.00 | 0.211 | 0.09 |
|  | L | **3.12** | 26.14 | 6.28 | 3.21 | 3.46 |
| Scientific | A | **0.16** | 0.92 | -0.55 | -0.015 | -0.24 |
|  | C | **0.72** | 0.02 | 0.00 | 0.002 | 0.00 |
|  | L | **6.83** | 18.01 | 9.18 | 4.74 | 8.17 |

The results about small world properties shown in Table 3 are for different statuses. There is no information on how the small world properties change from one status to another. Figure 8 shows the transition information from the original network topology to the four reshuffled statuses. The x-axis shows the change of degree correlations, while the plotted curves show small world properties measured by clustering coefficient and average path length. Every curve reflects a transition from the original network topology to the four reshuffled statuses: correlation+, correlation-, neutral, and random. Red curves denote variation of clustering coefficient *C*, while blue ones denote that of average path length *L*. The original network topology is indicated by two dots. The change of *C* and *L* from the initial network topology to both positive and negative degree correlations (in particular, toward correlation+) was more dramatic than the neutral and random statuses. The smooth (rather than fluctuated) curves indicate that both *C* and *L* are highly related to the variation of *A*. The assortativity coefficient *A* of all street network topologies (with the exception of Manhattan) is very close to zero. This makes them seem like the random status. However, the original network topologies are closer to small worlds than the random ones. This implies that topologies with a similar assortativity coefficient may have distinct structural differences.

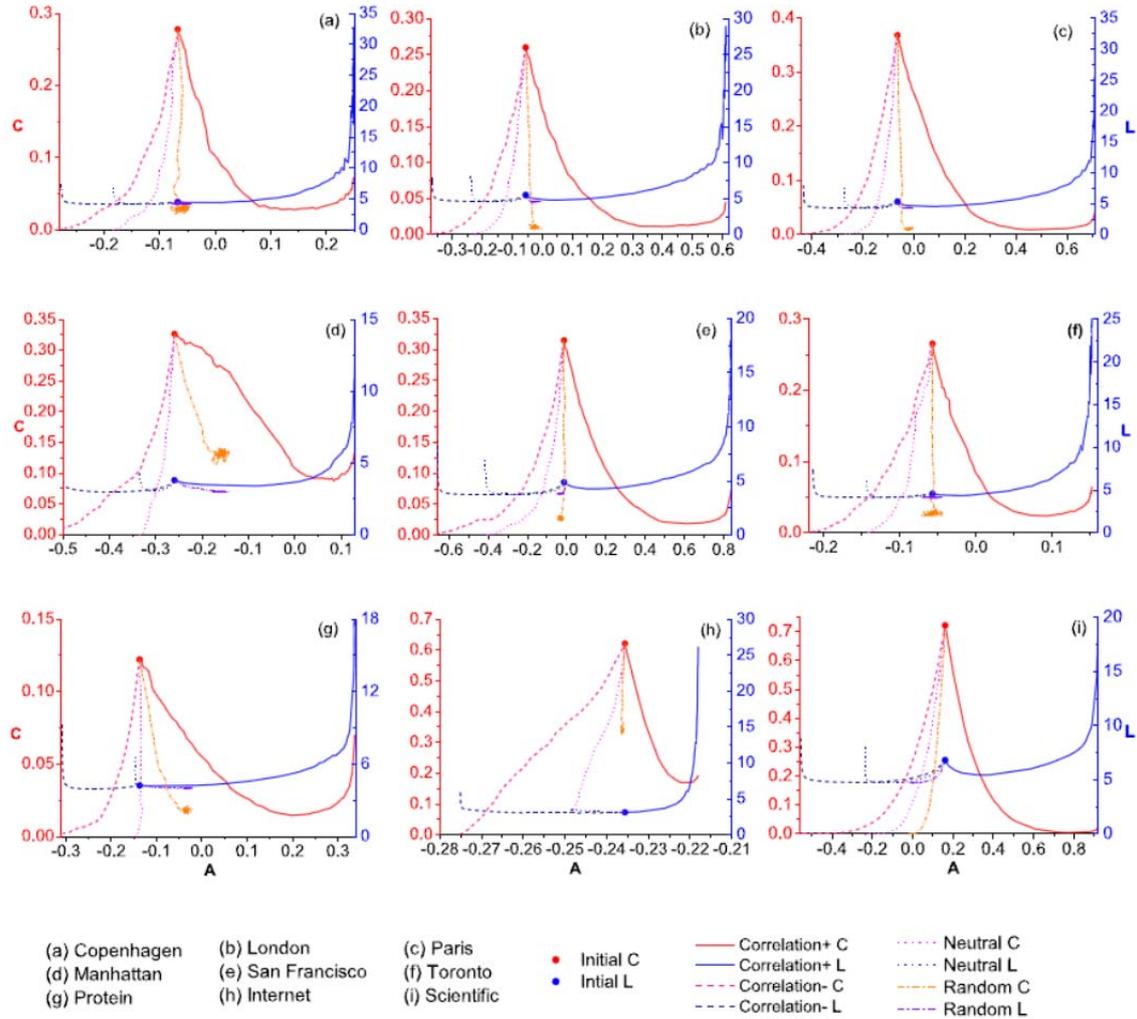

Figure 8: (Color online) Network efficiency with respect to the five statuses of the network topologies

## 4. Related work

One of the major findings of the study about the relationship between degree correlations and small world properties can be stated in another way. That is, scale free networks with high degree correlations are not necessary to be small world networks. This is consistent with the major finding of Small et al. (2008), who proposed a growth algorithm to examine the issue. The growth algorithm begins with a small complete graph. Given a degree distribution function of $P(d) = d^{-\alpha}$ and assuming the degrees of a new node $d_0$, which must meet a certain condition, the newly added node will be linked to existing $d_0$ nodes. The new node is linked to the existing graph through randomly (shuffled) and sequentially (unshuffled) chosen fittest $d_0$ nodes (i.e., similar degrees are linked to similar degrees).

The growth algorithm is based on a greed principle, which suffers from the problem that locally optimal does not always lead to globally optimal. A new node is added in every step, following degree correlation (similar degrees of nodes linked to each other), but it does not guarantee the entire graph holds degree correlation. In other words, the newly added node is most degree correlated with its neighbors, but it has changed the degree correlations between its neighbors and the neighbors' neighbors. The graph may not be most degree correlated at a global level. In this regard, the reshuffling process adopted in the study must be converged to a stable value for assortativity coefficient, although there were two pairs of links for reshuffling for every step. The entire graph, based on the reshuffling process through repeated iterations, can obtain the most stable statuses. With the growth process, the path length increases as the graph increases in size, although the increasing speed of path length is far quicker than that of graph growth. Subsequently it is difficult to judge which factors contribute to the increase of path length.

In our study, the networks' sizes never change in the reshuffling process, so it is pretty sure that the reshuffling process increase path length (or decreases the network efficiency). In addition, both $C$ and $L$ provide evidence to support that small world properties disappear from the initial network topologies to the four other statuses (becoming non-small world); $C$ dropped dramatically in all nine cases. However, Small et al.'s (2008) study showed that only $L$ provided striking evidence. Furthermore, the $L$ evidence was only available for one of the models (the sequential or unshuffled model). This was probably due to the shortcoming of the growth process as indicated above.

Another related work is Holme and Zhao (2007) who constructed the ensemble of networks with the same degree sequence as the original network, but changing assortativity and clustering coefficients. Using several biological networks, they studied how assortativity and clustering coefficient influence small world properties and robustness in the context of the network ensemble. This paper used a similar algorithm to construct networks of different statuses of assortativity by reshuffling the links while keeping the degree distributions of the original network. Using these constructed networks, this paper made some interesting findings related to topological features of street networks, which differ significantly from other complex networks.

## 5. Conclusion

This paper characterized the topological structure of urban street networks based on degree correlations compared to other complex networks. The degree correlations are described by different measures and from both local and global perspectives. We showed that in general urban street network topologies lack degree correlations ; this is to exclude Manhattan-like case in particular. We provided a sound explanation as to why there is such a lack of degree correlations. This is because on the one hand highly connected streets tend to connect to highly connected streets, and on the other hand, they tend to connect to less connected streets as well. This dual aspect offsets the degree correlations. Furthermore, the lack of degree correlation for street network topologies is certainly different from random and neutral graphs. The street networks exhibit high efficiency. However, all surrogate networks of the original street networks, even with a slightly different assortative coefficient, exhibit a dramatic reduction in network efficiency, i.e., dramatically deviating from small world properties.

We further investigated the relationship between degree correlations and the small world properties, and found that the small world properties disappear dramatically when networks deviate from the initial status. This finding for all the nine networks indicates that degree correlations do not imply high efficiency or high performance of complex networks. To better understand or interpret this finding, we tend to believe that complex networks of all kinds are self-organized or self-evolved. There is little room for improving their efficiency or performance. Such an improvement must involve their own self-organization processes of different complex networks rather than artificial reshuffling. This is rather different from the re-wiring process starting from a regular graph (Watts and Strogatz 1998), since a few re-wired links would lead to a small world dramatically. Our investigation also concluded that a scale-free network does not imply a small-world network. Although the topic has been studied in the literature, our study provides better and robust investigational settings. Future work can be extended to better measures for characterizing degree correlations or applications of the research for city planning and design.

**Acknowledgement**
We would like to thank Shi Zhou for kindly sharing the three reference networks studied in this paper. Jing Zhao acknowledges the financial support from the National Natural Science Foundation of China (10971227).